\documentclass[aps,showpacs,11pt,nofootinbib]{revtex4}
\usepackage{dcolumn}
\usepackage{graphicx}
\usepackage{amsmath}
\usepackage{amsfonts}
\usepackage{amssymb}
\usepackage{psfrag}
\usepackage{wrapfig}
\usepackage{subfigure}
\usepackage{makeidx}
\usepackage{bm}
\usepackage{epsf}
\usepackage{hyperref}
\usepackage{color}
\usepackage{multirow}

\begin{document}

\title{Magnetic-induced Spontaneous Scalarization in Dynamcial Chern-Simons Gravity}

\author{Shao-Jun Zhang$^{1,2}$}
\email{sjzhang@zjut.edu.cn}
\author{Bin Wang$^{3,4}$}
\email{wang$_$b@sjtu.edu.cn}
\author{Eleftherios Papantonopoulos$^{5}$}
\email{lpapa@central.ntua.gr}
\author{Anzhong Wang$^{6}$}
\email{Anzhong$_$Wang@baylor.edu}
\affiliation{$^1$Institute for Theoretical Physics $\&$ Cosmology, Zhejiang University of Technology, Hangzhou 310032, China\\
	$^2$United Center for Gravitational Wave Physics, Zhejiang University of Technology, Hangzhou 310032, China\\
	$^3$Center for Gravitation and Cosmology, College of Physical Science and Technology, Yangzhou University, Yangzhou 225009, China\\ $^4$Shanghai Frontier Science Center for Gravitational Wave Detection, Shanghai Jiao Tong University, Shanghai 200240, China\\
	$^5$Physics Department, National Technical University of Athens, 15780 Zografou Campus, Athens, Greece\\
	$^6$GCAP-CASPER, Physics Department, Baylor University, Waco, Texas 76798-7316, USA}

\date{\today}

\begin{abstract}
	\indent In the framework of the dynamical Chern-Simons gravity, we study the scalar field perturbations of the Reissner-Nordstr\"{o}m-Melvin spacetime, which describes a charged black hole permeated by a uniform magnetic field. In the presence of the magnetic field, the scalar field acquires an effective mass whose square takes negative value in the half domain of the angular direction. This  inevitably introduces the tachyonic instability and associated spontaneous scalarization as long as the coupling constant between the scalar field and the Chern-Simons invariant exceeds a threshold value. We study the object pictures of the time evolutions of the scalar field perturbations at the linear level, and find that the presence of the magnetic field will dramatically change the waveforms and associated ringdown modes. Nonlinear evolutions for the unstable perturbations are also performed in the decoupling limit, which demonstrate the scalar cloud as the final fate. Influences of the coupling constant and the black hole charge on the wave dynamics are also studied.
\end{abstract}


\maketitle

\section{Introduction}

In recent years there is a lot of activity trying to understand the plethora of  astrophysical observations. First of all, the LIGO and Virgo collaborations reported the detection of Gravitational Waves (GW)  from a Binary Black Hole (BH) Merger \cite{LIGOScientific:2016aoc,LIGOScientific:2016sjg,LIGOScientific:2017bnn,LIGOScientific:2017ycc,LIGOScientific:2020iuh}. More recently the First M87 Event Horizon Telescope reviled the shadow of a supermassive BH \cite{EventHorizonTelescope:2019dse}. Also, the recent results of the  Event Horizon Telescope are very interesting  detecting a magnetic field structure near the event horizon of a BH \cite{EventHorizonTelescope:2021srq}. General Relativity (GR) describes accurately the formation and the properties of BHs. A BH is the end state of the collapse of a very massive   cloud \cite{Tolman}-\cite{Penrose}. The generated structure is characterized by only three parameters, the mass,  charge and  angular momentum,  obeying the powerful no-hair theorems \cite{Bekenstein:1998aw,Robinson:2004zz,Chrusciel:2012jk}.

In  the same time many astrophysical observations showed that  many strong radio sources take the form of two emitting regions situated on opposite sites of a galaxy. A theory to account for this is that the magnetic fields and high energy particles responsible for the synchrotron radiation were blown out of the galactic halo  in a giant explosion.  It was speculated that such explosions could be generated from gravitational collapse. A simple model was presented \cite{Melvin:1963qx}, which was composed by a configuration that contains only electromagnetic field in a form of a collection of parallel magnetic lines hold together by its own gravitational attraction. Again GR can describe such a theory as a rigorous static cylindrically-symmetric solution of the combined sourceless Einstein-Maxwell equations \cite{Melvin:1965}-\cite{Thorne2} [For another possibility, see \cite{opher1996geodesic,Bronnikov:2019clf} and references therein].

The recent observational results on dark matter and on dark energy require  a generalization of GR in an attempt to have a viable cosmological theory of Gravity on short and large distances \cite{Joyce:2014kja,Nojiri:2006ri,Clifton:2011jh,Will:2014kxa,MG3}. We expect that these modified gravity theories (MOGs) will provide important information on the properties and structure  of the compact objects which are described by these theories and they are consistent with  the observational signatures, which they  introduce. In these theories matter sources parameterize by a  scalar field is directly coupled to second order algebraic curvature invariants. One of the most well studied MOGs theories  are the scalar-tensor theories in which matter is interacting with black holes like the  scalar-tensor Horndeski theories \cite{Horndeski}. Exploring the strong field regime
of gravity with the aim to detect  gravitational waves and black hole shadows the effects of higher-order curvature terms become  significant. The most well known high dimensional theories are the scalar-Einstein-Gauss-Bonnet theory (sEGB) \cite{Torrii_1996}-\cite{Guo:2020zqm} and the dynamical Chern-Simons gravity theory (dCSG) \cite{Cardoso:2009pk}-\cite{Chatzifotis:2022ene}.

In dSCG, a dynamical scalar field is introduced to couple to the gravitational Chern-Simons invariant non-minimally \cite{Jackiw:2003pm}. The motivation for studying dSCG is that it is a gravity theory that contain higher powers of curvature which are consistent
Lanczos-Lovelock gravity theories in higher than four dimensions. The Chern-Simons black holes are special solutions of these
theories resulting in second order field equations for the metric with well defined AdS asymptotic solutions.  The action for CS modified gravity is defined by the sum of the Einstein-Hilbert action and a
new parity violating, four-dimensional correction.  Interest in the model spiked when it was found that string
theory unavoidably requires such a correction to remain mathematically consistent \cite{Campbell:1990fu,Moura:2006pz}. In general, such a correction arises in the presence of Ramond-Ramond scalars because of the presence of duality symmetries and these parity violating
scalar fields can dress a slowly rotating black hole with  scalar hair.

One of the first work of a hairy   slowly rotating black hole, which resulted because  the scalar field was coupled to a Lorentz CS term, was presented in \cite{kerr1,Campbell:1991kz}. This result suggested  that non-minimal gravitational couplings may produce interesting new effects in black hole backgrounds. These works were  further extended in \cite{kerr3} where it was found that the  scalar hair are characterized by the mass, angular momentum and gauge charges of the background rotating black hole. Allowing the scalar field to have a dynamical behavior a solution describing a rotating black hole  in the small coupling slow rotation limit was found in \cite{Yunes:2009hc}. Static and rotating black string solutions  in dynamical CS modified gravity were also studied in \cite{Cisterna:2018jsx,Corral:2021tww}.

The coupling of a scalar field to  high curvature terms allows us to evade  the no-hair theorems  and obtained more general hairy black hole configurations. In particular, for certain classes of the coupling function it was shown in the case of GB high curvature terms, that we have spontaneous scalarization of black holes \cite{Doneva_2018a}-\cite{Antoniou_2018a}. It was found in regions
of strong curvature, that below a n critical mass the Schwarzschild black hole develops instabilities   and then when the scalar field backreacts to the metric,   new branches of scalarized black holes are generated at certain masses as solutions in the  theory \cite{Doneva_2018a,Silva_2018,Myung_2018b}.  Actually, this novel phenomenon has been observed long time ago in neutron stars but there the instability is triggered by the surrounding matter instead of the curvature \cite{Damour:1993hw}. The mechanism of spontaneous scalarization can also be applied to the Chern-Simons theory. More precisely, on the Kerr BH background, the massless scalar field perturbations will acquire an effective mass due to its coupling to the CS invariant. The square of the effective mass becomes negative in half domain of angular directions resulting in the tachyonic instability, and thus leading to the  spontaneous scalarization.

However, most of the studies on spontaneous scalarization  have considered that the background BHs live in vacuum.  In the realistic universe, BHs are usually not isolated but always live in complicated astrophysical environments, such as magnetic fields, accretion disks, dark matter halos, cosmological expansion, etc. These environments may have considerable influences on dynamics of BHs under perturbations, dramatically change the waveforms and associated ringdown modes \cite{Barausse:2014tra,Brito:2014nja}. Then one can ask  how much of the impact does the realistic astrophysical environments have on the phenomenon of spontaneous scalarization?

In this work,  we will focus on the influences of one factor of these environments, the magnetic field. It is believed to be ubiquitous in nature and pervade our universe on various scales with different amplitudes, ranging from order of $\sim 10^{-4}$ Gauss at the center of our galaxy \cite{Crocker:2010xc} to order of $\sim 10^{16}$ Gauss at surface of some magnetars \cite{Olausen:2013bpa}. Also, existences of BHs permeated in magnetic backgrounds are also supported by astrophysical observations of recent years. For example, the presence of the magnetar SGR J$1745$-$2900$ orbiting the supermassive BH Sagittarius $A^\ast$ \cite{Mori:2013yda,Kennea:2013dfa,Eatough:2013nva,Olausen:2013bpa}, and the presence of strong magnetic field in vicinity of the event horizon of $M 87^\ast$ \cite{EventHorizonTelescope:2021srq}. Moreover, it is believed that strong magnetic fields around BH play a central role in some of the most energetic events of our universe, such as emission of relativistic jets via, for example, the Blandford-Zanjek process \cite{Blandford:1977ds}.

In general, full comprehension of the interactions between BH and the surrounding magnetic field is a rather involved problem. As a simple model, we consider stationary magnetized BH solutions in which a qualitative picture can be drawn. Actually,  a class of exact solutions of the Einstein-Maxwell equations has been known, describing BHs immersed in a uniform magnetic field aligned along the symmetry axis, the Kerr-Newmann-Melvin (KNM) BHs \cite{Wald:1974np,Ernst:1976mzr,Ernst-Wild:1976,Gibbons:2013yq}. For a recent review of this subject, we refer readers  to \cite{Budinova:2000yd,Bicak:2006hs}. Different from the well-known Kerr-Newmann BH in GR, the KNM BH spactime is not asymptotically flat but resembles the magnetic Melvin universe \cite{Melvin:1963qx}. The thermodynamical properties of these BHs have been studied extensively in recent years \cite{Gibbons:2013dna,Astorino:2016hls,Booth:2015nwa,Astorino:2015naa,Astorino:2016ybm}.

In the KNM model, effects of the magnetic field on the spontaneous scalarization is studied preliminary in sEGB theory in \cite{Annulli:2022ivr,Hod:2022qir}.
With this  by analyzing the behavior of effective mass square of the scalar field perturbation near the horizon directly, it was found  that the presence of the magnetic field pushes up the threshold dimensionless spin of the BH for spin-induced tachyonic instability. However, a complete object picture of the times evolution of the scalar field perturbations to show the influences of the magnetic field on the dynamics is still lack. Interestingly, it is found in \cite{Brihaye:2021jop} that even in the absence of compact objects (like BHs), a real massive scalar field can still condense in the Melvin magnetic universe when coupled non-minimally either to the magnetic field or to the curvature. Another type of mechanism generating BH hairs - the superradiant instability in these BHs has also been studied in \cite{Brito:2014nja,Santos:2021nbf}. By considering massless neutral scalar field perturbations, they found that the presence of magnetic fields enacts both requirements simultaneously for the occurrence of the superradiant instability and formation of stationary scalar clouds, namely an ergoregion to trigger superradiance and a confinement mechanism to trap the scalar field. In \cite{Soldateschi:2020hju} a magnetized compact object, like a neutron star, was studied in scalar-tensor theories.

Inspired by these works, we will study the effects of magnetic fields on spontaneous scalarization in dCSG. We take  Reissner-Nordstrom-Melvin (RNM) BH as the background. This BH is characterized by three parameters, the BH mass $M$, the BH charge $Q$ and the magnetic field $B$. It reduces to the RN BH in the limit $B \rightarrow 0$, while reduces to the Schwarzschild-Melvin solution in the limit $Q \rightarrow 0$ \cite{Ernst:1976mzr}. It should be noted that in the absence of magnetic fields, the CS invariant (and thus the effective mass square that the scalar field acquires) vanishes identically. So in this case, there is no tachyonic instability and spontaneous scalarization.

However, when the magnetic field is present, situation changes in two aspects. On the one hand, the scalar field acquires an effective mass whose square may become negative thus would trigger tachyonic instability. On the other hand, the presence of magnetic fields will change the asymptotical structure of the spactime which then may influence the waveforms of the scalar field perturbations. Therefore the presence of the magnetic field introduces high symmetries making the system more involved compared to a system of  astrophysical black holes which are nearly asymptotically flat. Then natural questions arise: How do the presence of the magnetic field affect the stability of the BH? And how do the presence of the magnetic field influence the wave dynamics of scalar perturbations? These are main questions we would like to answer in this work.

The paper is organized as follows. In Sec. II, we will give a brief introduction of dCSG theory and the RNM BH and establish the scalar field perturbation equation. In Sec. III, we describe our numerical method to solve the scalar field perturbation equation in detail. In Sec. IV, we report our numerical results. The final section is devoted to summary and discussions.

\section{The model}
In this work, we utilize the units $c=G=4\pi \epsilon_0 =1$, where $c, G, \epsilon_0$ are the the speed of light in vacuum, the Newton gravitational constant and the vacuum permittivity respectively.

We consider dCSG with an additional electromagnetic field. The general action is \cite{Alexander:2009tp}
\begin{eqnarray}
	S&=& \frac{1}{16 \pi}\int d^4x \sqrt{-g}\left[R - F_{\mu\nu}F^{\mu\nu}+\alpha h(\Phi) \ ^\ast R R+{\cal L}_{\Phi}\right],\nonumber\\
	{\cal L}_\Phi&=&-\frac{1}{2}\nabla^\mu\Phi\nabla_\mu\Phi-V(\Phi),\nonumber
\end{eqnarray}
where the real scalar field $\Phi$ is non-minimally coupled to the CS invariant through the coupling constant $\alpha$ and the coupling function $h(\Phi)$. $V(\Phi)$ is the scalar self-interaction potential. The explicit definition of the CS invariant is
\begin{eqnarray}
	^\ast R R \equiv \frac{1}{2} \epsilon^{\alpha \beta \gamma \delta} R_{\alpha \beta \mu}^{~~~~\nu} R_{\gamma \delta \nu}^{~~~~\mu}.
\end{eqnarray}
By varying the action, one can derive the equations of motion
\begin{eqnarray}
	\nabla^2\Phi&=&\frac{dV}{d\Phi}-\alpha  ~^\ast R R \frac{d h}{d\Phi},\label{ScalarEq}\\
	\nabla_\mu F^{\mu \nu} &=& 0, \label{GaugeEq}\\
	R_{\mu\nu}-\frac{1}{2}g_{\mu\nu}R &=&\alpha T^{CS}_{\mu\nu}+T^\Phi_{\mu\nu} + T^{EM}_{\mu\nu},\label{MetricEq}\\
	T^{CS}_{\mu\nu}&=& -4 \nabla^\sigma h \epsilon_{\sigma \alpha \beta (\mu} \nabla^\beta R_{\nu)}^{~\alpha} - 2 \nabla^\alpha \nabla^\beta h \ \epsilon_{\alpha \rho \sigma (\mu} R_{\nu)\beta}^{~~~~\rho\sigma},\nonumber\\
	T^\Phi_{\mu\nu}&=&\frac{1}{2}\nabla_\mu\Phi\nabla_\nu\Phi-\frac{1}{2} g_{\mu\nu}V(\Phi)-\frac{1}{4} g_{\mu\nu}\nabla^\rho\Phi\nabla_\rho \Phi.\nonumber\\
	T^{EM}_{\mu\nu}&=& 2 F_{\mu \rho} F_\nu^{~\rho} - \frac{1}{2} g_{\mu \nu} F_{\rho \sigma} F^{\rho \sigma}.  \nonumber
\end{eqnarray}

In this work, we consider the scalar field to be massless without self-interaction so that $V(\Phi)=0$, and take the form of the coupling function as
\begin{eqnarray}
	h(\Phi) = \frac{1}{2 \beta} \left(1 - e^{-\beta \Phi^2}\right),
\end{eqnarray}
where $\beta$ is a constant. With these choices, the theory admits GR electro-vacuum solutions with vanishing scalar hair  $\Phi= 0$, among which to our interest in this work is the Reissner-Nordstr\"{o}m-Melvin (RNM) solution \cite{Ernst:1976mzr,Gibbons:2013yq}
\begin{eqnarray}
	ds^2 &=& H \left[-f dt^2 + f^{-1} dr^2 + r^2 d\theta^2\right] + H^{-1} r^2 \sin^2\theta(d\varphi - \Omega dt)^2, \label{RNMelvin}\\
	A_\mu dx^\mu &=& \phi_0 dt + \phi_3 (d\varphi - \Omega dt),
\end{eqnarray}
where
\begin{eqnarray}
	f &=& 1 - \frac{2 M}{r} + \frac{Q^2}{r^2},\nonumber\\
	H &=& 1 + \frac{1}{2} B^2 \left(r^2 \sin^2\theta + 3 Q^2 \cos^2\theta\right) + \frac{1}{16} B^4 \left(r^2 \sin^2\theta + Q^2 \cos^2\theta\right)^2,\nonumber\\
	\Omega &=& -\frac{2 Q B}{r} + \frac{Q B^3 r}{2} \left(1 + f \cos^2\theta\right),\nonumber\\
	\phi_0 &=& -\frac{Q}{r} + \frac{3}{4} Q B^2 r \left(1 + f \cos^2\theta\right),\nonumber\\
	\phi_3 &=& \frac{2}{B} - H^{-1} \left[\frac{2}{B} + \frac{B}{2} \left(r^2 \sin^2\theta + 3 Q^2 \cos^2\theta\right)\right],
\end{eqnarray}
which describes a RN black hole with charge $Q$ permeated by a uniform magnetic field with strength $B$ aligned along the symmetry axis. Note that this BH solution has cylindrical symmetry \cite{Bronnikov:2019clf}, and is not asymptotically flat but resembles the magnetic Melvin Universe \cite{Melvin:1963qx}. The event horizon is located at $r=r_+ = M + \sqrt{M^2 - Q^2}$. It reduces to the RN black hole solution in the limit $B \rightarrow 0$, while reduces to the Schwarzschild-Melvin solution  in the limit $Q \rightarrow 0$ \cite{Ernst:1976mzr}. On this background, we would like to study the wave dynamics of scalar field perturbations and study possible instabilities. Then the scalar field perturbation equation (\ref{ScalarEq}) becomes
\begin{eqnarray}
	\nabla^2\Phi=m^2_{\rm eff} \Phi, \label{ScalarEq1}
\end{eqnarray}
\begin{figure}[!htbp]
	\includegraphics[width=0.45\textwidth]{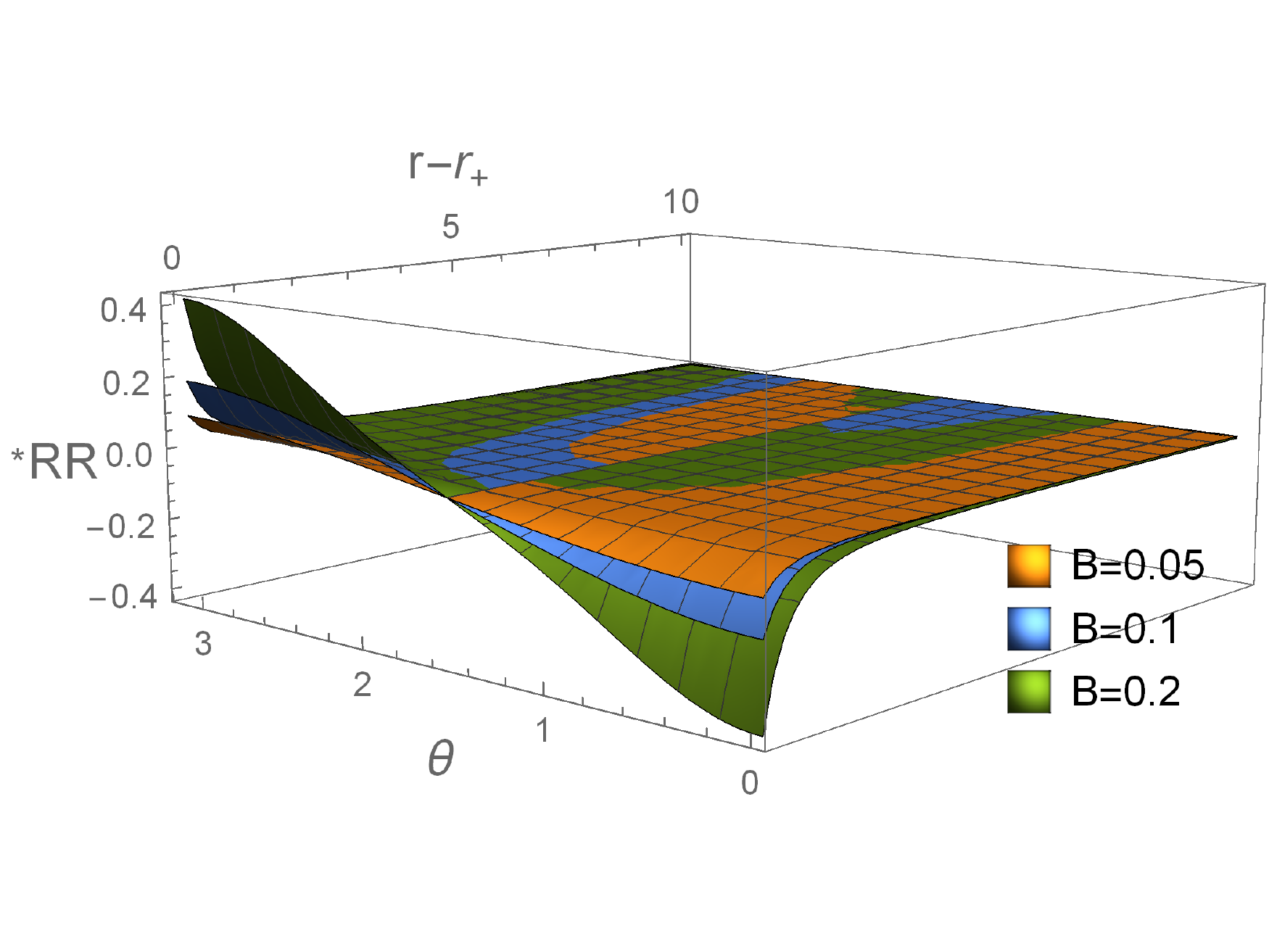}\quad
	\includegraphics[width=0.45\textwidth]{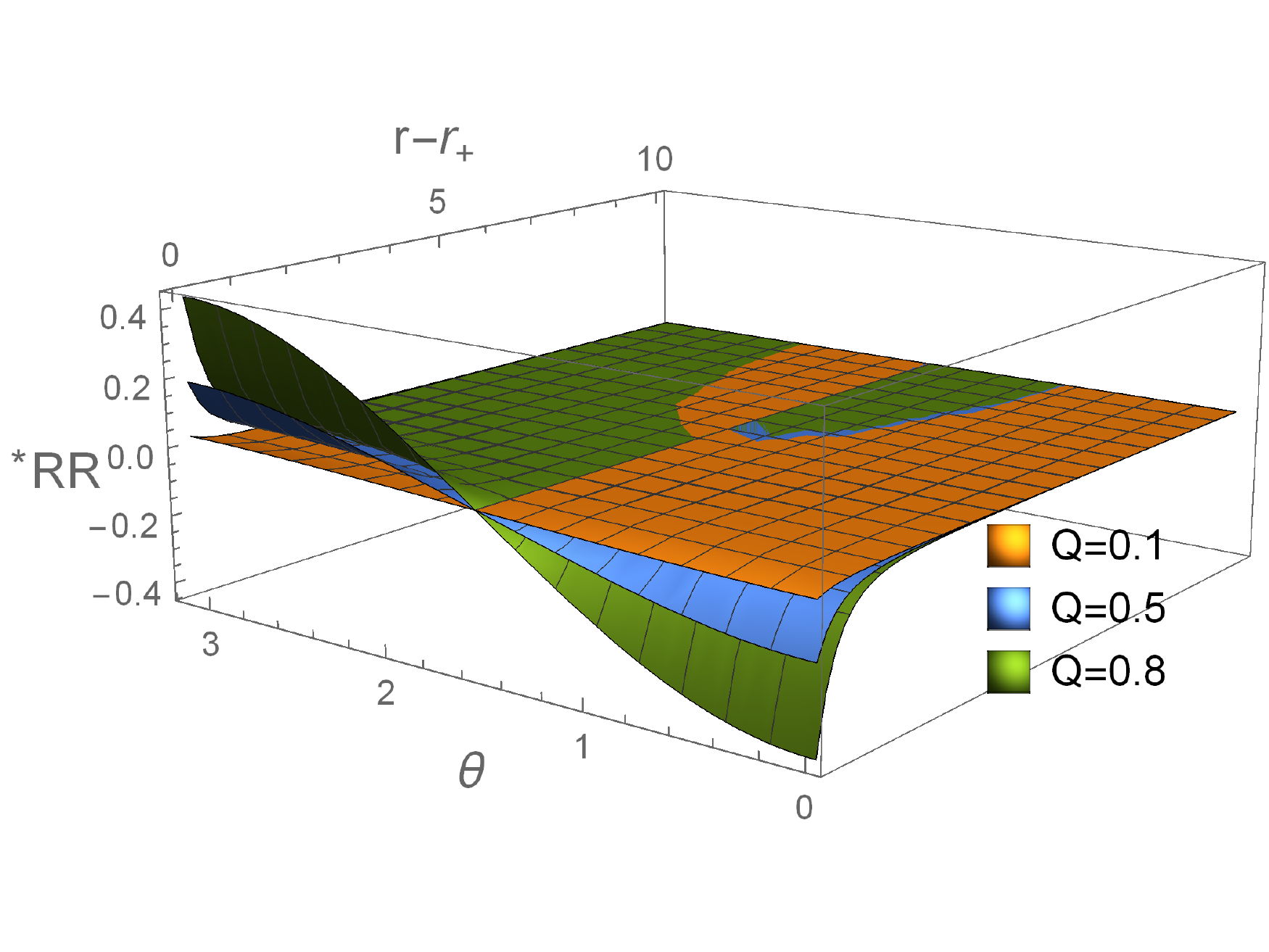}
	\caption{(color online) The CS invariant $^{\ast}RR$ as a function of $r$ and $\theta$. In the left panel, $Q=0.5$ is fixed while $B$ is varied; In the right panel, $B=0.1$ is fixed while $Q$ is varied.} \label{CSFig}
\end{figure}
where the effective mass square $m^2_{\rm eff}=-\alpha ~^\ast R R$ is position-dependent and depends on parameters $(M, Q, B)$. The  full explicit expression of the CS invariant $^\ast R R$, valued in the background, is rather involved and uninspiring and will not be displayed here.  One can get a preliminary feeling of the influence of the magnetic field on the CS invariant from its small-$B$ expansion
\begin{align}
	^\ast R R = & \frac{96 Q \cos \theta \left(Q^2-M r\right)}{r^6} B -\frac{12 Q \cos\theta}{r^6} \left[Q^2 r (7 r-6 M)\right.\nonumber             \\
	            & \left.+\cos 2 \theta \left(-Q^2 r (32 M+23 r)+3 r^3 (4 M+r)+32 Q^4\right)+r^3 (r-8 M)+6 Q^4\right] B^3 +{\cal O} \left(B^5\right).
\end{align}
It should be noted that throughout this work we will perform the computations with the full form of the CS invariant rather than its small-$B$ expansion.
Several comments on the scalar field perturbation equation (\ref{ScalarEq1}) now are in order:
\begin{itemize}
	\item For vanishing magnetic field $B=0$ (vanishing BH charge $Q=0$), the background reduces to RN (Schwarzschild-Melvin) BH, and the CS invariant and the effective mass square vanish identically. In this case, the above equation describes wave propagation of a massless neutral scalar field in the RN (Schwarzschild-Melvin) background which has already been studied thoroughly and no instability is observed \cite{Brito:2014nja,Berti:2009kk,Konoplya:2011qq}.
	\item The CS invariant is odd under the transformation
	      \begin{eqnarray}
		      \theta \rightarrow \pi-\theta \qquad \Rightarrow  \qquad {}^\ast RR \rightarrow - {}^\ast RR. \label{parity}
	      \end{eqnarray}
	      So in the scalar field perturbation equation (\ref{ScalarEq1}), the sign of $\alpha$ can always be absorbed into the CS invariant by redefining the $\theta$-coordinate. Taking into account this symmetry, we will only consider $\alpha>0$ in the following. Moreover, it also means that, for non-zero magnetic field $B$/BH charge $Q$, the effective mass square $m^2_{\rm eff}$ will always take negative value in half interval of  $\theta$ domain, implying the inevitable tachyonic instability as long as the coupling constant $\alpha$ is large enough.
\end{itemize}
In Fig. \ref{CSFig}, profiles of the CS invariant are shown for some typical values of parameters. From the figure, one can see that the CS invariant $^\ast RR$ takes a finite non-zero value near the horizon while approaches to zero at infinity. This can be understood because the CS term is a high curvature term, so  the strong gravitational field is expected to occur only near the horizon of the BH while it is negligible far away from the BH. Moreover, increasing the magnetic field $B$/BH charge $Q$ will make the CS invariant more positive (thus the effective mass square $m^2_{\rm eff}$ more negative for $\alpha>0$) in the half interval of $\theta$ domain near the horizon, hence triggering more violent instability.

In the following sections, we will study carefully the time evolution of the massless scalar field perturbations and obtain object pictures of the influences of the magnetic field $B$, the coupling constant $\alpha$ and also the BH charge $Q$ on wave dynamics.

\section{Numerical method}

We will apply the numerical method discussed in \cite{Krivan:1996da,PazosAvalos:2004rp,Dolan:2011dx,Doneva:2020nbb,Schiesser} to solve the scalar field perturbation equation (\ref{ScalarEq1}).
With the RNM metric Eq. (\ref{RNMelvin}), the scalar field perturbation equation (\ref{ScalarEq1}) becomes
\begin{eqnarray}
	&&r^2 \partial_t^2 \Phi- f \partial_r \left(f r^2 \partial_r \Phi\right) + 2 r^2 \Omega \partial_t \partial_\varphi \Phi\nonumber\\
	&&- \frac{f}{\sin\theta} \partial_\theta \left(\sin\theta \partial_\theta \Phi\right) + \left(r^2 \Omega^2 -\frac{f H^2}{\sin^2 \theta}\right) \partial_\varphi^2 \Phi= \alpha ~^\ast R R f H r^2 \Phi.
\end{eqnarray}
After introducing the tortoise coordinate $x$ defined as $dx \equiv dr/f$ to map the radial domain $r \in (r_+, +\infty)$ to $x \in (-\infty, +\infty)$, the above equation becomes
\begin{eqnarray}
	&&r^2 \left(\partial_t^2 - \partial_x^2\right) \Phi- 2 f r \partial_x \Phi+ 2 r^2 \Omega \partial_t \partial_\varphi \Phi\nonumber\\
	&&- \frac{f}{\sin\theta} \partial_\theta \left(\sin\theta \partial_\theta \Phi\right) + \left(r^2 \Omega^2 -\frac{f H^2}{\sin^2 \theta}\right) \partial_\varphi^2 \Phi= \alpha ~^\ast R R f H r^2 \Phi.
\end{eqnarray}
In the case of scalar field perturbations in the Kerr BH \cite{Krivan:1996da}, it has already been noted that the azimuthal coordinate $\varphi$ will introduce unphysical pathologies near the horizon, which can be removed by adopting the Kerr azimuthal coordinate \cite{Krivan:1996da}. In the current case, a similar problem will appear and can be solved by introducing a Kerr-like azimuthal coordinate $\tilde{\varphi}$ defined as
\begin{eqnarray}
	d \tilde{\varphi} = d \varphi + \Omega dx.
\end{eqnarray}
Then the equation becomes
\begin{eqnarray}
	&&r^2 \left(\partial_t^2 - \partial_x^2\right) \Phi- 2 f r \partial_x \Phi+ 2 r^2 \Omega \left(\partial_t \partial_{\tilde{\varphi}} - \partial_x \partial_{\tilde{\varphi}}\right) \Phi \nonumber\\
	&&- \partial_x (r^2 \Omega) \partial_{\tilde{\varphi}} \Phi - \frac{f}{\sin\theta} \partial_\theta \left(\sin\theta \partial_\theta \Phi\right) -\frac{f H^2}{\sin^2 \theta} \partial_{\tilde{\varphi}}^2 \Phi  = \alpha ~^\ast R R f H r^2 \Phi.
\end{eqnarray}
Due to the axial symmetry of the RNM spacetime, the scalar field perturbation can be decomposed as
\begin{eqnarray}
	\Phi (t, x, \theta, \tilde{\varphi}) = \sum_m \Psi (t, x, \theta) e^{i m \tilde{\varphi}},
\end{eqnarray}
with $m$ being the azimuthal number. Then, by introducing an auxiliary variable $\Pi \equiv \partial_t \Psi$, the perturbation equation can be cast into a form of two coupled first-order partial differential equations
\begin{eqnarray}
	\partial_t \Psi = &&\Pi,\nonumber\\
	\partial_t \Pi = &&- 2 i m \Omega \Pi + \partial_x^2 \Psi + 2 \left(i m \Omega + \frac{f}{r}\right) \partial_x \Psi + \frac{f}{r^2} \left(\partial_\theta^2 + \cot\theta \partial_\theta\right)\Psi \nonumber\\
	&&+\left(\frac{i m}{r^2} \partial_x (r^2 \Omega) - \frac{m^2 f H^2}{r^2 \sin^2\theta} + \alpha ^\ast RR f H\right) \Psi,\label{ScalarEqFinal}
\end{eqnarray}
which are suitable for the method presented in \cite{Schiesser}. Precisely, we apply the fourth-order Runnge-Kutta integrator to perform the time evolution, while the finite difference scheme for the spatial derivatives.

To solve the perturbation equation, we also need to impose physical boundary conditions. At the horizon, ingoing wave condition is implemented following \cite{Ruoff2000}. At the outer boundary, the magnetic field will behave like an infinite ``wall" at $x \sim 1/B$ to confine the perturbation \cite{Brito:2014nja}, so it is sensible to impose a Dirichlet boundary condition $\Psi = 0$ there. Also, at the poles $\theta=0$ and $\pi$, we should impose physical boundary conditions $\Psi|_{\theta=0, \pi} = 0$ for $m \neq 0$ while $\partial_\theta \Psi|_{\theta=0, \pi} = 0$ for $m=0$ \cite{Dolan:2011dx}.

\section{Numerical results}

We consider the scalar field perturbation initially to be a Gaussian wave-packet localized outside the horizon at $x=x_c$ which has time symmetry \cite{Zhang:2020pko,Zhang:2021btn},
\begin{eqnarray}
	\Psi (t=0, x, \theta) &\sim& Y_{\ell m} e^{-\frac{(x - x_c)^2}{2 \sigma^2}},\\
	\Pi (t=0, x, \theta) &=& 0,
\end{eqnarray}
where $Y_{\ell m}$ is the $\theta$-dependent part of the spherical harmonic function and $\sigma$ is the width of the Gaussian wave-packet.

As in the Kerr spacetime, the RNM spacetime we consider here is not spherically symmetric except when $B=0$, so mode-mixing phenomenon \cite{Thuestad:2017ngu,Zenginoglu:2012us,Burko:2013bra} will also occur during evolution: a pure initial $\ell$-multipole
will excite other $\ell'$-multipoles with the same $m$ as it evolves with time. Taking into account this phenomenon and for simplicity, in the following we will only consider axisymmetric perturbations with $\ell = m = 0$. Also, we set $M = 1$ so that all quantities are measured in
units of $M$. Without loss of generality, observers are assumed to locate at $x = 6 M$ and $\theta = \frac{\pi}{5}$.

\subsubsection{Object pictures: Linear and nonlinear level}

\begin{figure}[!htbp]
	\includegraphics[width=0.45\textwidth]{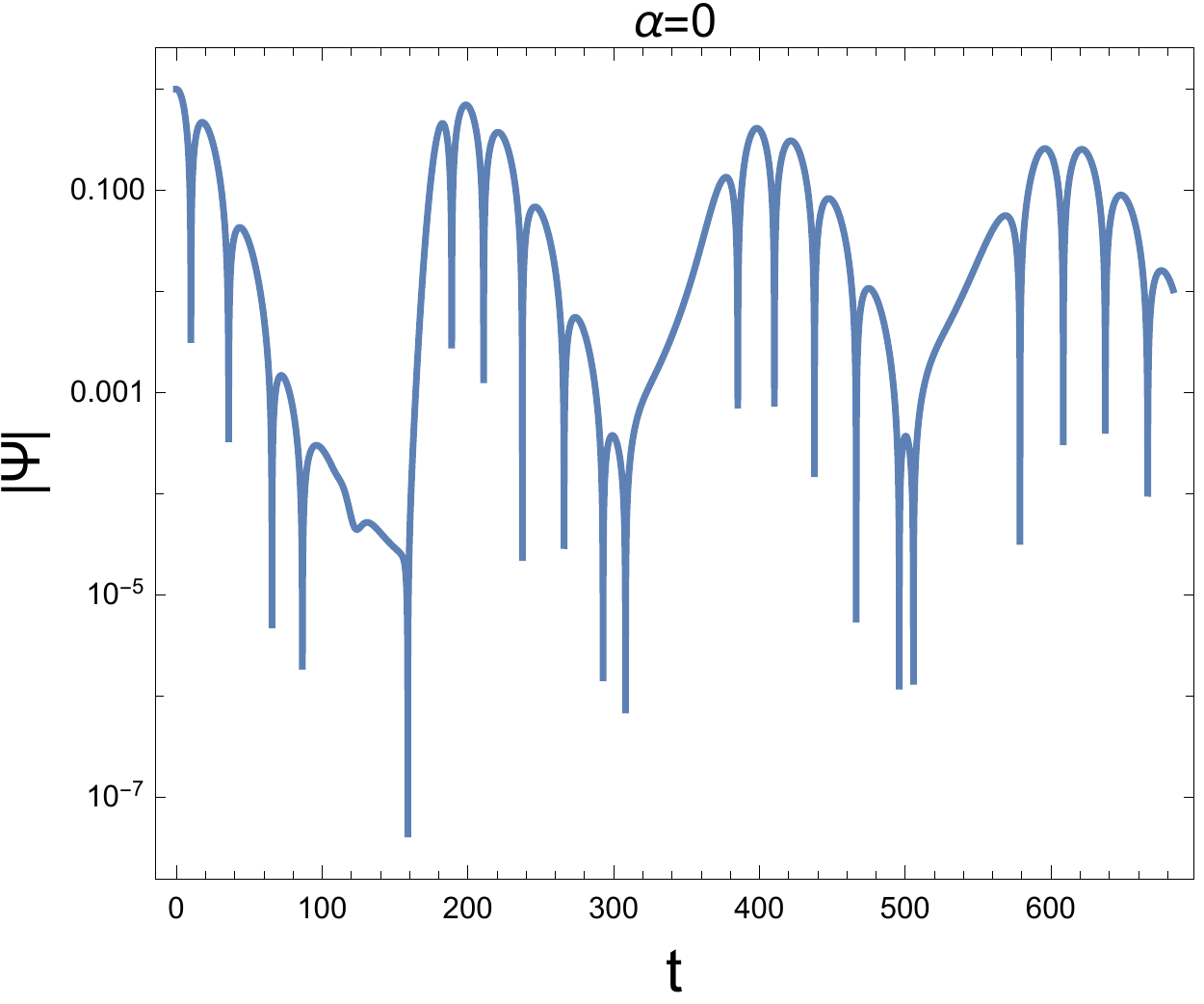}
	\includegraphics[width=0.45\textwidth]{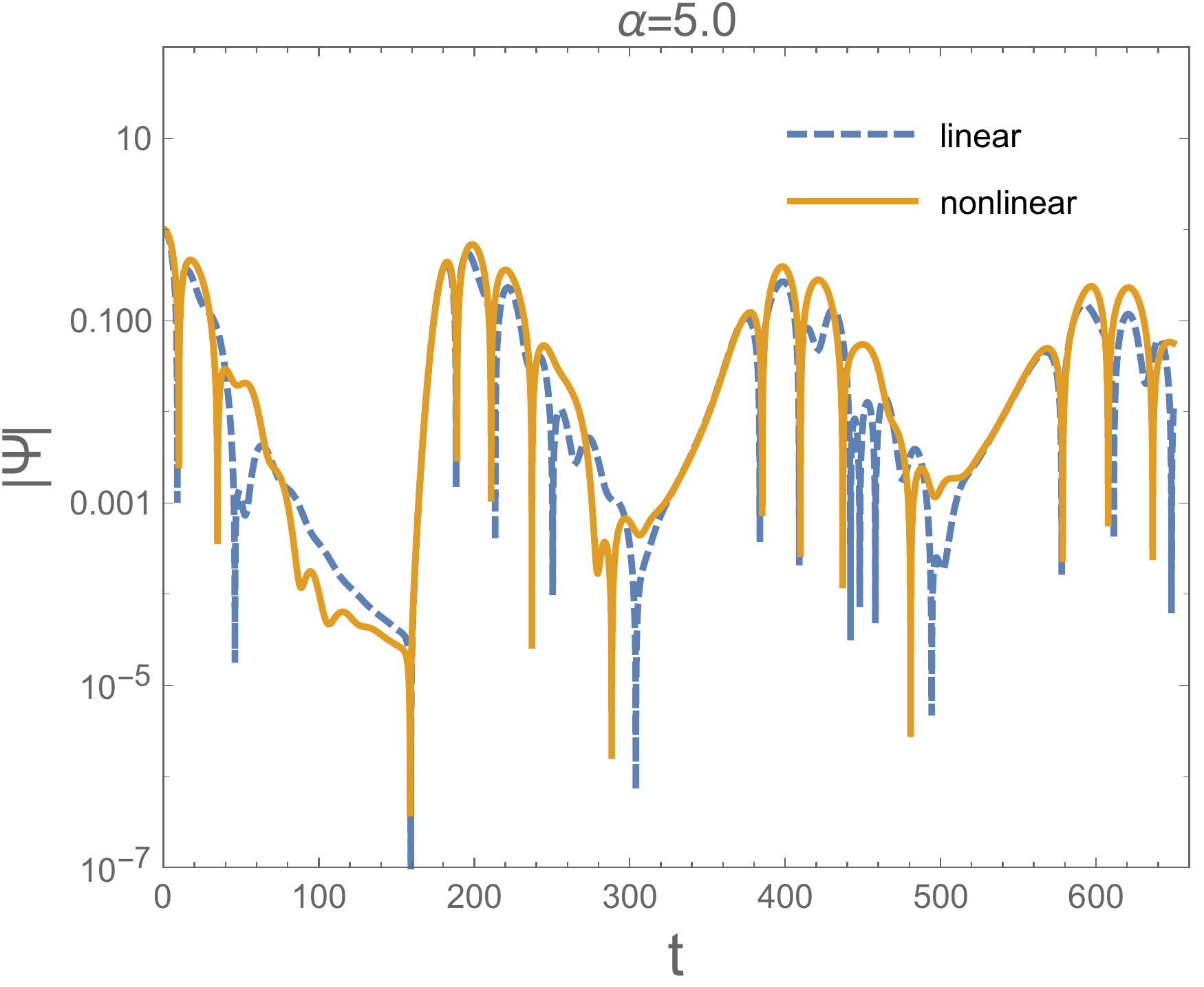}
	\includegraphics[width=0.45\textwidth]{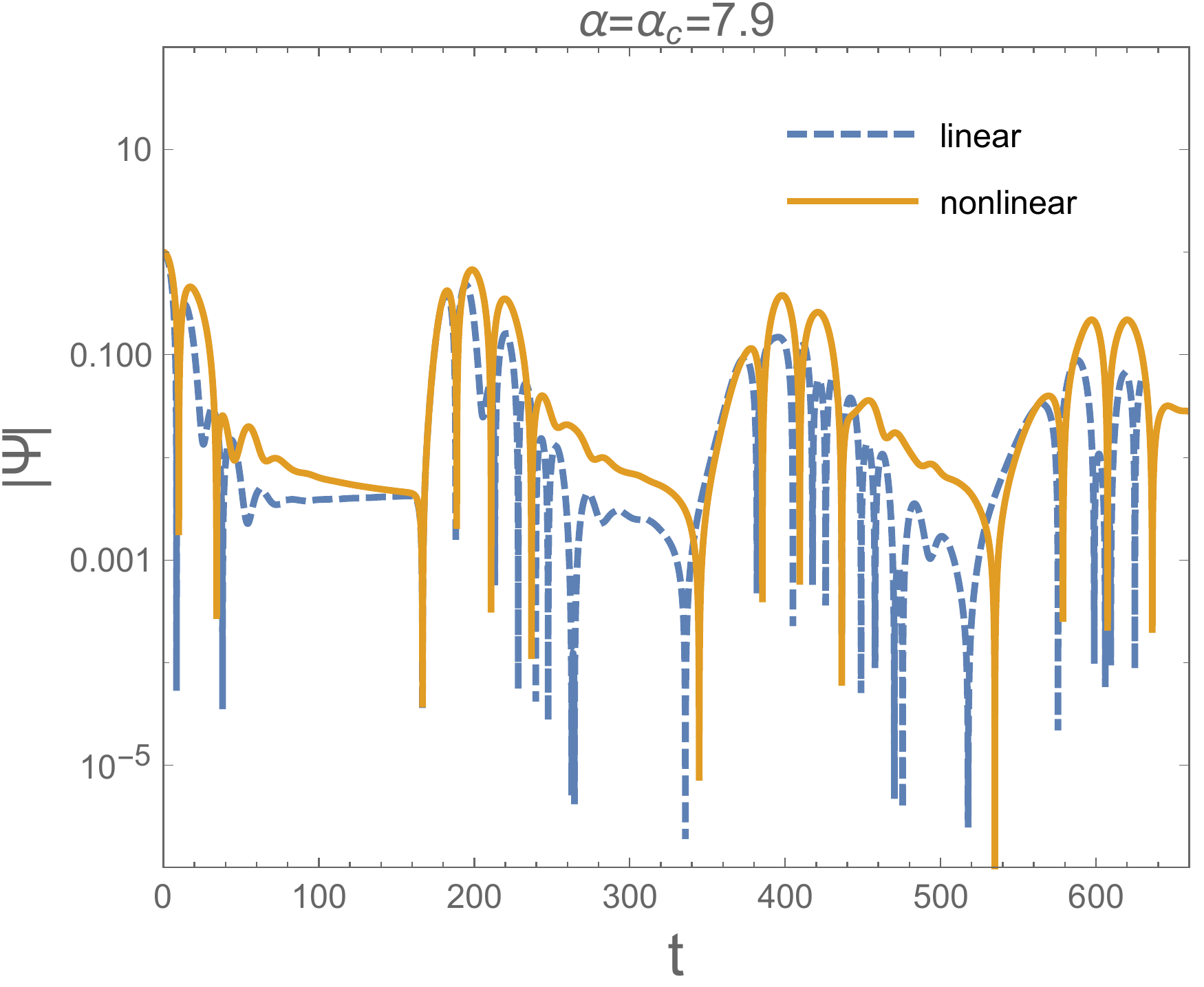}
	\includegraphics[width=0.45\textwidth]{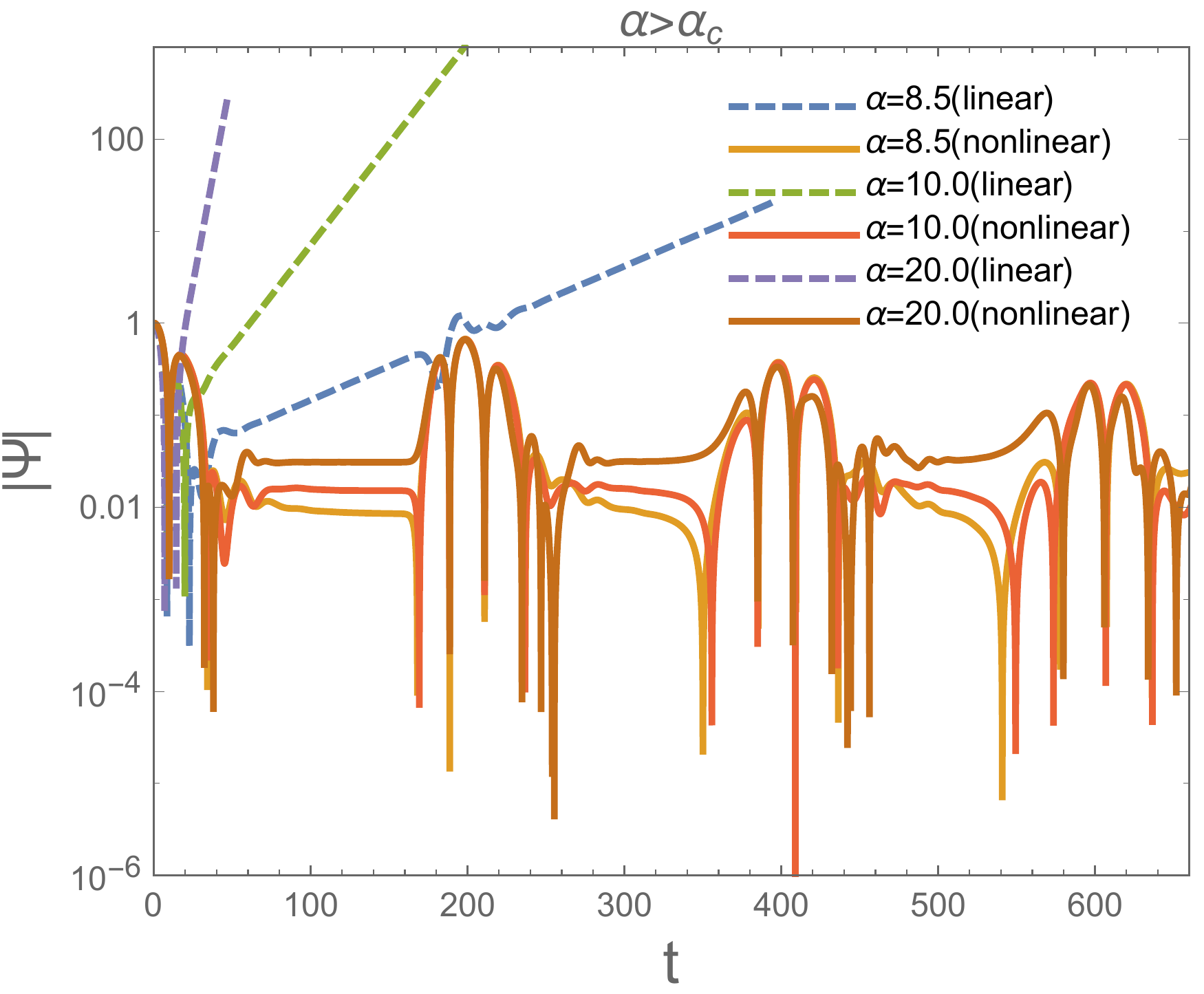}
	\caption{(color online) Time evolutions of the scalar field perturbation for  $B=0.1, Q=0.5$ and various values of $\alpha$ both in the linear and nonlinear levels (with the nonlinear parameter $\beta=10.0$). From left to right and top to bottom, $\alpha$ increases from $0$ to $20.0$. Initial data of the perturbation is $\Psi (t=0, x) \sim e^{-\frac{(x - x_c)^2}{2 \sigma^2}}$ with $x_c=6.0$ and $\sigma=6.0$. } \label{ObjectPicture}
\end{figure}

In Fig. \ref{ObjectPicture}, time evolutions of the scalar field perturbations are shown for $B=0.1, Q=0.5$ and various values of the coupling constant $\alpha$ in linear level. For other values of parameters, we have similar phenomena. Several  properties are observed:
\begin{itemize}
	\item There exists a critical value of the coupling constant $\alpha=\alpha_c \simeq 7.9$, above which tachyonic instability occurs. Moreover, with the increase of $\alpha$, the instability occurs earlier and becomes more violent. This is expected as the effective mass square $m_{\rm eff}^2 \propto \alpha$.
	\item For $\alpha \leq \alpha_c$, the waveforms show two types of ringdown modes with different frequencies at different stages. This interesting phenomenon has already been observed in  \cite{Brito:2014nja}, where minimally coupled neutral scalar field perturbations of the Schwarzschild-Melvin BH in framework of GR is studied. Following the arguments there, in our case one can expect that at early times, the ringdown modes are being in fact similar to the quasinormal modes (QNMs)  of RN BH under  massive scalar field perturbations with the additional effective mass $\mu_{\rm eff} = m B$ \cite{Galtsov:1978ag,Konoplya:2007yy,Konoplya:2008hj}, which are excited near the horizon. After a time of order $t \sim \frac{1}{B}$, the wave is reflected back by the effective ``wall" induced by the magnetic field and the so-called ``Melvin-like" modes with smaller amplitudes are excited \cite{Brito:2014nja}. Both types of modes decay with time and are expected to vanish finally. See also  \cite{Barausse:2014tra} for more discussions on similar interesting phenomenon in the context of  ``dirty" BHs.
	\item When $\alpha = \alpha_c$, the waveform between the two types of modes approaches a constant indicating criticality.
	\item For $\alpha > \alpha_c$, unstable tachyonic modes are excited and the Melvin-like modes are greatly suppressed. When $\alpha$ is large enough, the Melvin-like modes are completely over-dominated by the unstable tachyonic modes and thus have not been observed.
\end{itemize}

To see the final fate of the instability, we have also studied the time evolutions of the scalar field perturbation in the nonlinear level but in the so-called ``decoupling limit" \cite{Doneva:2021dqn,Doneva:2021dcc}. That is, we take into account the non-linearities in the scalar field coupling while neglecting its back-reaction on the spacetime. This limit has been shown to be able to capture the basic features of the full nonlinear dynamics qualitatively. In this limit, one should consider the scalar field equation with the full coupling
\begin{eqnarray}
	\nabla^2\Phi=m^2_{\rm eff} \frac{d h(\Phi)}{d\Phi}.
\end{eqnarray}
Time evolutions of the scalar field perturbation in this limit are shown in Fig. \ref{ObjectPicture} for $\alpha \neq 0$. When $\alpha=0$, $m^2_{\rm eff}=0$ and there are no nonlinear effects. From the figure, one can see that when $\alpha=5.0 < \alpha_c$,  there is little difference between the linear and nonlinear levels. This is expected as nonlinear effects can be neglected for stable perturbations. However, for $\alpha > \alpha_c$, when taking into account the nonlinear effect, the instability is quenched and the unstable tachyonic modes are suppressed to approach a non-zero constant while the Melvin-like modes thus revive. With the slowly decaying of the Melvin-like modes in time, it is expected that the scalar field will approach a non-zero constant finally to form a scalar cloud around the BH. Moreover, it is also observed that with the increase of $\alpha$, the scalar field approaches a larger constant finally which can be understood physically.

\subsubsection{Effects of $B$ and $Q$}

\begin{figure}[!htbp]
	\includegraphics[width=0.45\textwidth]{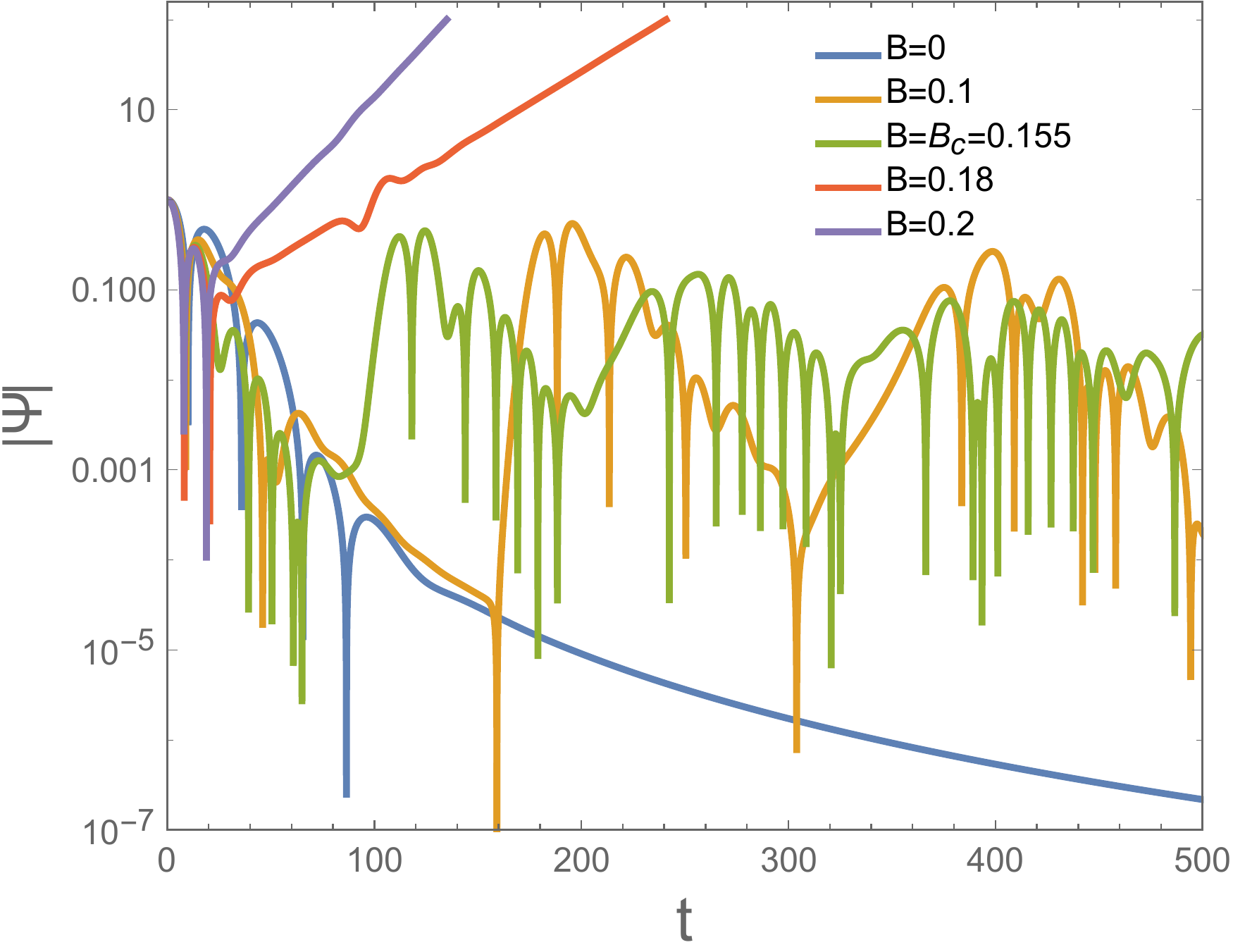}\quad
	\includegraphics[width=0.43\textwidth]{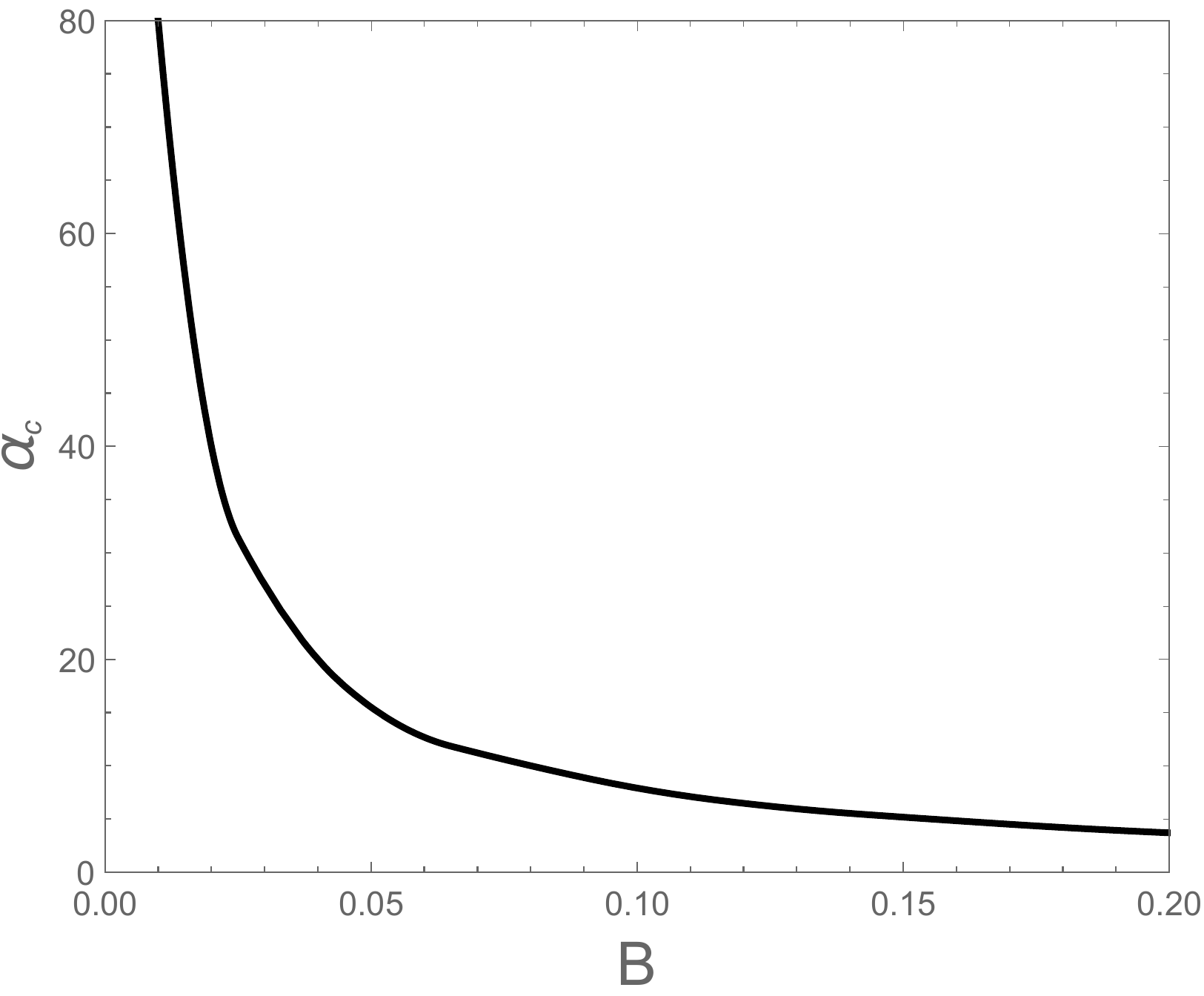}
	\caption{(color online) {\em Left}: time evolutions of the scalar field perturbation in the linear level for  $\alpha=5.0, Q=0.5$ and various values of $B$.  Initial data of the perturbation is $\Psi (t=0, x) \sim e^{-\frac{(x - x_c)^2}{2 \sigma^2}}$ with $x_c=6.0$ and $\sigma=6.0$. {\em Right}: critical value of the coupling constant $\alpha_c$ versus $B$ for fixed $Q=0.5$.} \label{EffectsB}
\end{figure}

\begin{figure}[!htbp]
	\includegraphics[width=0.45\textwidth]{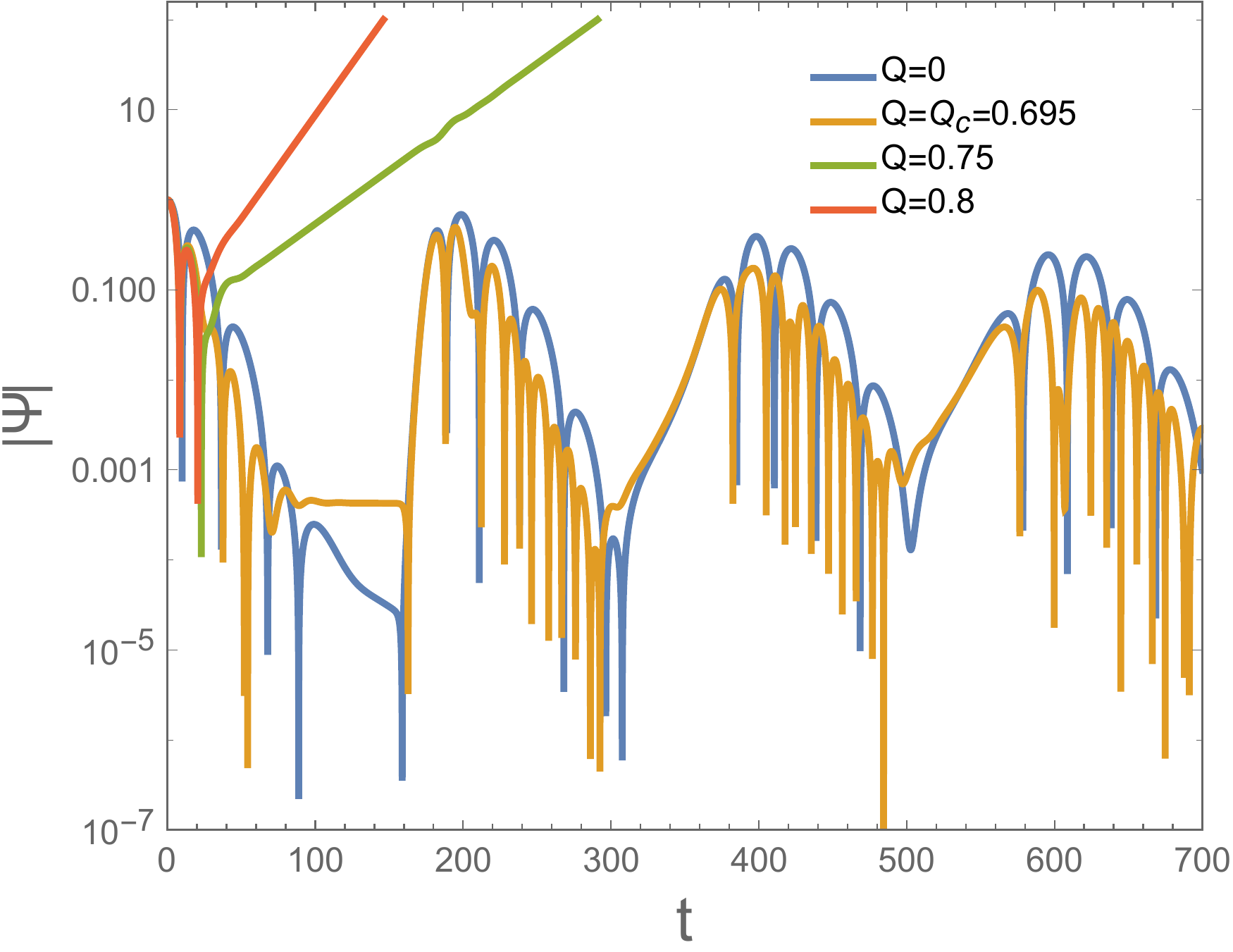}\quad
	\includegraphics[width=0.42\textwidth]{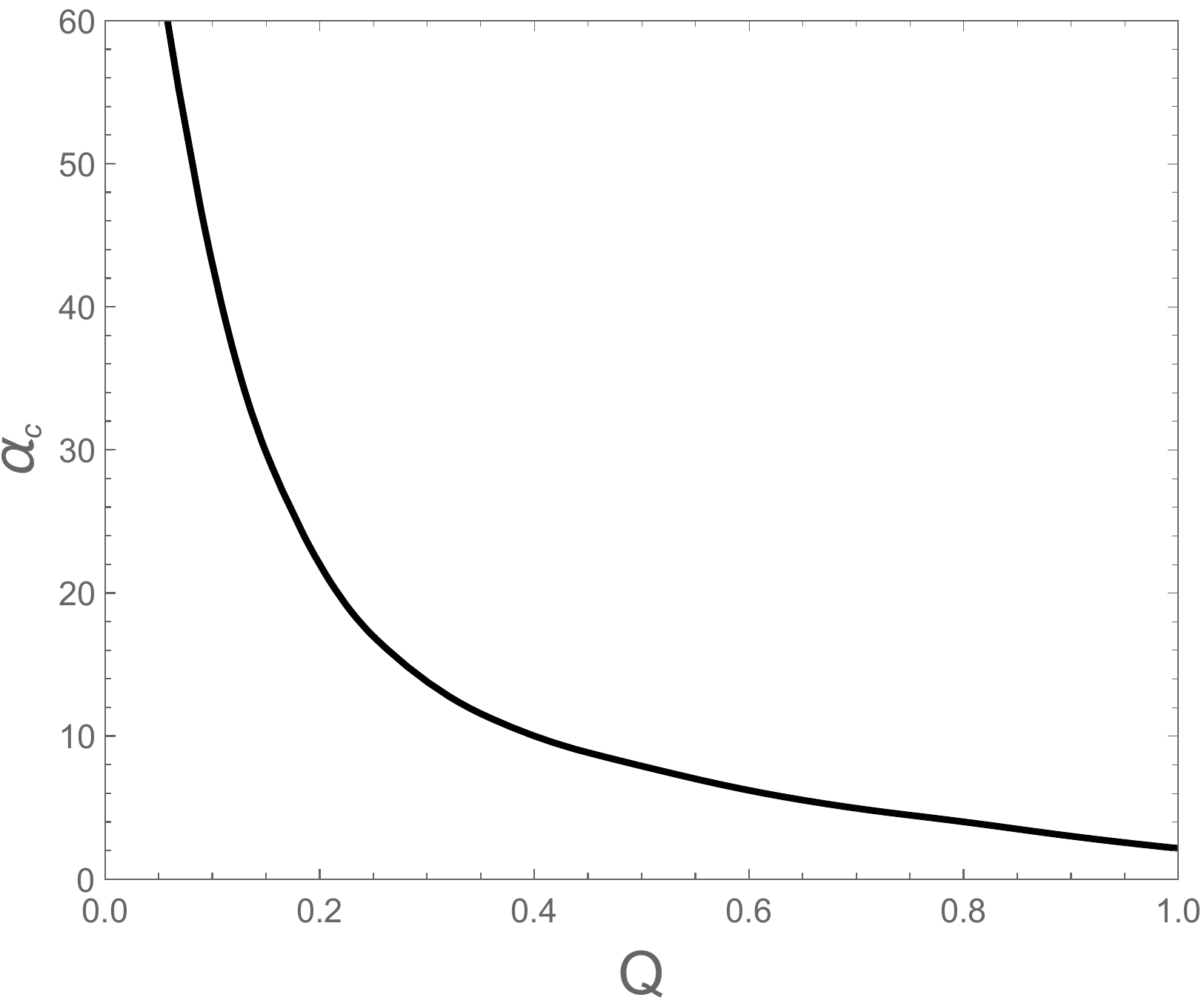}
	\caption{(color online) {\em Left}: time evolutions of the scalar field perturbation in the linear level for $\alpha=5.0, B=0.1$ and various values of $Q$. Initial data of the perturbation is $\Psi (t=0, x) \sim e^{-\frac{(x - x_c)^2}{2 \sigma^2}}$ with $x_c=6.0$ and $\sigma=6.0$. {\em Right}: critical value of the coupling constant $\alpha_c$ versus $Q$ for fixed $B=0.1$.} \label{EffectsQ}
\end{figure}

In Figs. \ref{EffectsB} and  \ref{EffectsQ}, we plot time evolutions of the scalar field perturbation for various values of $B$ and $Q$ to see their effects on the dynamics of wave propagation. Meanwhile,  we also show their influences on the critical value of the coupling constant $\alpha_c$. In these figures, parameters $(\alpha, B, Q)$ take some sample values. For other values, similar behaviors are observed.

From the left panel of Fig. \ref{EffectsB}, one can see that there exists a critical value of $B=B_c \simeq 0.155$, above which tachyonic instability is triggered. For $B \leq B_c$, increasing $B$ will make the effective ``wall" closer to the horizon and thus the Melvin-like modes excited earlier. For $B>B_c$, once again we can see that the Melvin-like modes is suppressed by the unstable tachyonic modes. Increasing $B$ will make the instability to occur earlier and more violent, which can be understood physically from Fig. \ref{CSFig} as larger $B$ will make the effective mass square more negative near the horizon. The case with $B=0$ is also shown for comparison, from which one can see that there is no Melvin-like modes but instead the usual decaying tail after the early ringdown stage. So the existence of even small $B$ will dramatically change the waveform. Comparing to Fig. \ref{ObjectPicture} and from the right panel of Fig. \ref{EffectsB}, one can see that the critical value of the coupling constant $\alpha_c$ decreases with the increase of $B$.

From the left panel of Fig. \ref{EffectsQ}, one can see that the effect of $Q$ on the dynamics of the wave propagation is similar to that of $B$. Also, there exists a critical value of $Q=Q_c \simeq 0.695$ above which tachyonic instability is triggered. However, the biggest difference from the effects of $B$ is that even $Q=0$, there are still Melvin-like modes and the waveform does not change much for small $Q$. This is expected as the existence of $B$ will change the asymptotical structure of the spacetime dramatically while $Q$ will not. Also, by comparing to Fig. \ref{ObjectPicture} and from the right panel of Fig. \ref{EffectsQ}, one can see that the critical value of the coupling constant $\alpha_c$ also decreases with the increase of $Q$.

\section{Summary and Discussions}

In this work, we studied  the scalar field perturbations of RNM BH in the framework of dCSG theory. Due to the coupling to the CS invariant, the scalar field acquires an effective mass whose square $m^2_{\rm eff}$ will always take negative value in the half interval of $\theta$ domain, thus implying the inevitable tachyonic instability as long as the coupling constant $\alpha$ is large enough. Object pictures of the wave dynamics of the scalar field perturbations in time for various values of parameters are obtained as shown in Figs. \ref{ObjectPicture}, \ref{EffectsB} and \ref{EffectsQ}, which confirm our conclusions. Note that the CS invariant ${}^{\ast}RR$ (and thus the effective mass square) vanish when $B=0$, so this tachyonic instability and associated spontaneous scalarizations are magnetic-induced. The critical value of the coupling constant $\alpha_c$ to trigger instability decreases with the increase of $B$ or $Q$. And larger $B$ or $Q$ make the instability to occur earlier and more violent. These phenomena can be easily understood physically from the behavior of the CS invariant as shown in Fig. \ref{CSFig}.

From the figures, one can see more interesting phenomena induced by the magnetic field $B$. As the existence of $B$ changes dramatically the asymptotical structures of the spacetime, waveforms of the scalar field perturbation are very different from that of $B=0$. When $\alpha \leq \alpha_c$, two types of different ringdown modes with different frequencies appear at different stages. At early time, the ringdown modes are similar to QNMs of RN BHs under a massive scalar field perturbation with additional effective mass $\mu_{\rm eff} = m B$ \cite{Galtsov:1978ag,Konoplya:2007yy,Konoplya:2008hj}. While after a time of order $t \sim \frac{1}{B}$, the so-called ``Melvin-like" modes with smaller amplitude are excited due to the reflections of the wave from the effective ``wall" induced by the magnetic field \cite{Brito:2014nja}. Both types of modes decay with time (although very slowly) and are expected to vanish finally. However, when $\alpha > \alpha_c$, unstable tachyonic modes are excited and the Melvin-like modes are suppressed greatly. When taking into account the non-linearities of the scalar field coupling in decoupling limit, the instability is quenched and the unstable tachyonic mode approaches to a non-zero constant to form a scalar cloud. It is expected that when back-reaction of the scalar field to the spacetime geometry is considered, the scalar cloud will become the scalar hair of the BH.

Let us give some comments on the astrophysical relevance of the parametric value of the magnetic field we considered. In this work, typical value of the magnetic field considered is $B M \sim 0.1$. Retaining physical units, we have the following relation
\begin{equation}
	\frac{1}{M} \simeq 2.36 \times 10^{19} \left(\frac{M_\odot}{M}\right) \textrm{Gauss},
\end{equation}
where $M_\odot$ is the solar mass. So $B M \sim 0.1$ corresponds to a magnetic field $B \sim 2.36 \times 10^{18} \left(\frac{M_\odot}{M}\right) \textrm{Gauss}$ whose explicit value is inversely proportional to the mass of the BH; For stellar-mass BHs with $M \sim 10 M_\odot$, $B \sim 10^{17}\textrm{Gauss}$, while for supermassive BHs with $M \sim 10^6 M_\odot$ (for example the Sagittarius A$^\ast$), $B \sim 10^{12} \textrm{Gauss}$. In general, for BHs with mass $M >10^2 M_\odot$, the magnetic field considered will be smaller than the ever-measured strongest magnetic field $B \sim 10^{16} \textrm{Gauss}$ \cite{Olausen:2013bpa}. Moreover, as have been observed, even for smaller magnetic field, the same tachyonic instability and spontaneous scalarization can still be triggered as long as the coupling constant $\alpha$ is large enough.

Of course, a more precise description of the formation of the hair requires nonlinear evolutions of the full system including the spacetime, which will be rather involved technically. It will also be interesting to construct the final spontaneously scalarized black hole solutions, which is still an open question in dCSG theory, whether in our considered model in this work or in the original model without magnetic field. Moreover, it will also be interesting to extend current studies to the rotating case by considering the background BH to be the Kerr-Melvin BH, or other models of spontaneous scalarizations, like the Einstein-Maxwell-scalar model
\cite{Herdeiro:2018wub}, to see the effects of the environmental magnetic field. We would like to leave these questions for further investigations.

\begin{acknowledgments}

	This work is supported in part by the National Natural Science Foundation of China (NNSFC) under Grant Nos. 11975203, 12075202, 12075207.
\end{acknowledgments}

\end{document}